\documentclass[12pt]{iopart}

\usepackage[bottom]{footmisc}

\usepackage{etoolbox}

\makeatletter
  \patchcmd{\@makefntext}{\fnsymbol}{\arabic}{}{}%
  \patchcmd{\@thefnmark}{\fnsymbol}{\arabic}{}{}%
  \def\@makefnmark{\textsuperscript{\arabic{footnote}}}
\makeatother
\usepackage{graphicx}

\usepackage{cite}
\usepackage{placeins}
\usepackage{iopams}
\usepackage{float}
\usepackage[breaklinks=true,colorlinks=true,linkcolor=blue,urlcolor=blue,citecolor=blue]{hyperref}

\usepackage{marginnote}

\begin{document}

\title{Entanglement Entropy of Disjoint Spacetime Intervals in Causal Set Theory}

\author{Callum F. Duffy, Joshua Y. L. Jones, and Yasaman K. Yazdi}
\address{Theoretical Physics Group, Blackett Laboratory, Imperial College London, %

SW7 2AZ, UK}
\ead{callum.duffy17@imperial.ac.uk, joshua.jones17@imperial.ac.uk, ykouchek@imperial.ac.uk}
\vspace{10pt}
%\begin{indented}
%\item[]August 2017
%\end{indented}

\begin{abstract}
A more complete understanding of entanglement entropy in a covariant manner could inform the search for quantum gravity. We build on work in this direction by extending previous results to disjoint regions in $1+1$D. We investigate the entanglement entropy of a scalar field in disjoint intervals within the causal set framework, using the spacetime commutator and correlator, \(i\mathbf{\Delta}\) and \(\mathbf{W}\) (or the Pauli-Jordan and Wightman functions). A new truncation scheme for disjoint causal diamonds is presented, which follows from the single diamond truncation scheme. We investigate setups including two and three disjoint causal diamonds, as well as a single causal diamond that shares a boundary with a larger global causal diamond. In all the cases that we study, our results agree with the expected area laws. In addition, we study the mutual information in the two disjoint diamond setup. The ease of our calculations indicate our methods to be a useful tool for numerically studying such systems. We end with a discussion of some of the strengths and future applications of the spacetime formulation we use in our entanglement entropy computations, both in causal set theory and in the continuum.

%A more complete understanding of entanglement entropy in a covariant manner could inform the search for quantum gravity. We investigate entanglement entropy for disjoint regions within the causal set framework, using the spacetime correlators \(i\mathbf{\Delta}\) and \(\mathbf{W}\) the Pauli-Jordan and Wightman function respectively. A new truncation scheme of the Pauli-Jordan function was found to be necessary, in order to recover continuum results for disjoint regions. This truncation scheme was verified by the use of simulations, whereupon we confirmed four entropy scalings from conformal field theory literature. The configurations investigated consisted of a single subregion on a boundary, two disjoint subregions, and three disjoint subregions. 

% The first result confirmed was the entanglement entropy scaling for a single causal diamond at the boundary of a larger diamond. We obtained a logarithmic scaling coefficient of \(0.165 \pm 0.0195\), which is in good agreement with the expected \(\frac{1}{6}\). We also calculated the entropy scalings with cutoff for two disjoint and three disjoint causal diamonds, located within a larger diamond. The coefficients on the logarithmic scalings obtained were \(0.669 \pm 0.0207\) and \(0.998 \pm 0.0681\) respectively, which are in good agreement with the expected \(\frac{2}{3}\) and \(1\). The mutual information decay of two disjoint diamonds as they become far separated was also confirmed.

\end{abstract}

\section{Introduction}

Entanglement entropy has often been considered a pathway to quantum gravity, ever since its introduction by Rafael Sorkin \cite{Sorkin1983}, where a  connection to black hole entropy was shown. In that work, the entanglement entropy of a quantum scalar field in the exterior of a black hole was shown to scale as the spatial area \(A\) of the event horizon, in units of the UV cutoff, \(A/\ell_{UV}^2\).  Likewise, the entropy of a black hole was known to scale as the area of the event horizon \cite{PhysRevD.7.2333}, leading to the proposal that the Bekenstein-Hawking entropy may be wholly or in large part entanglement entropy. Much subsequent research has investigated potential origins of black hole entropy \cite{Mathur_2005, Horowitz:1996qd, Rovelli_1996, Carlip_2015, Dou:2003af}, including entanglement entropy as a leading candidate \cite{PhysRevD.34.373, Jacobson_2013, Solodukhin_2011, Emparan_2006}. Presently, it remains an important open question as to what the true microscopic source of black hole entropy is.

In addition to its role in quantum gravity, entanglement entropy has proven to be a rich topic with many applications in  other areas of physics. A number of interesting theorems have been derived using entanglement entropy in quantum field theory, and especially in conformal field theory. These include c-theorems which relate a function of the coupling constants of the theory to the Virasoro central charge \cite{Casini_2004, Calabrese_2004, Myers_2010}. 
%state that there exists a positive real function of the coupling constants of a quantum field theory which is non-increasing along the RG trajectories and stationary at the fixed points, where it takes a finite value proportional to the Virasoro central charge (ref).
In AdS/CFT, Ryu and Takayanagi found that the entanglement entropy is equal to $1/4$ the area of a minimal surface \cite{Ryu_2006, Ryu2_2006}. %, and it has since been proven first in special cases (ref), and later in general (ref). 
In condensed matter physics, features such as topological order and properties of Fermi surfaces have been studied using entanglement entropy \cite{Kitaev_2006, PhysRevLett.96.110405, Swingle_2010}. In quantum information, entanglement entropy is used to constrain processes such as teleportation, and to study the properties of states \cite{Vidal_2002}. %Also, features of entanglement entropy are used in the study of Firewalls which are related to the black hole information paradox (ref). 

Just as there are many different applications of entanglement entropy, there are many different methods for computing it. These include heat kernel methods \cite{Solodukhin_2011}, the replica trick \cite{Callan_1994, Hertzberg_2012}, the Ryu-Takayanagi formula \cite{Ryu_2006}, Euclidean path integral methods \cite{Rosenhaus_2014, Rosenhaus_2015}%(especially for perturbative treatments of interacting theories)
, using spatial correlators \cite{PhysRevD.34.373, Peschel_2003}, and using spacetime correlators \cite{Sorkin:2012sn, Chen_2020}. The spacetime correlator method is what we use in this paper. The choice of technique used depends both on the physical features and requirements of the theory, as well as what is practically solvable. In this paper we are interested in the entanglement entropy of a Gaussian scalar field in disjoint $1+1$D spacetime intervals (causal diamonds) in a causal set. Causal sets do not admit a meaningful notion of a state on a hypersurface, which is clear upon consideration of a maximal antichain\footnote{An antichain is a subset of the causal set consisting of causally unrelated elements. A \emph{maximal} antichain is one that is inextendable, i.e. one cannot add more elements to the set such that they all remain causally unrelated.}, the analogue of a spatial hypersurface. Due to the random spatio-temporal discreteness of the causal set, causal connections can pass through the antichain without making an imprint on it. This means that the maximal antichain cannot carry all the information associated with spacetime elements to its past and future, and therefore cannot serve as a Cauchy surface. This renders unusable many of the conventional spatial methods for computing entanglement entropy. 

The method for computing entanglement entropy from spacetime correlation functions of the field, which we work with, was first introduced in \cite{Sorkin:2012sn}. We will review this formulation in Section \ref{sec:SfromW}. In addition to the requirement from causal set theory to work with spacetime rather than spatial quantities, there are also more general physical motivations to do so. An ultraviolet cutoff is required to obtain a finite and well-defined entanglement entropy, and if the entropy is computed on a spatial hypersurface, the cutoffs used are spatial and therefore not covariant. Ambiguities can then arise in making a choice of frame and cutoff, which are physically undesirable, especially in the context of gravity. The entanglement entropy formulation we work with has the advantage of admitting the use of a covariant spacetime UV cutoff.

In Section \ref{sec:oneint} we review the single interval entanglement entropy in a causal set, and we present new results for the case where the interval shares a boundary with the global region. We then discuss the multiple interval case in Sections \ref{sec:2int} and \ref{sec:3int}. We consider two and three disjoint regions, though our work easily generalises to any number of disjoint intervals. Treating disjoint intervals turns out to be another strength of working with the spacetime correlator formulation of entanglement entropy. This is because the details of the multiple interval calculation follow very closely  those of the single interval case, which has been well studied in \cite{Afshordi_2012, Saravani_2014, Sorkin_2018}, and only additionally requires the use of greater computational power.

\section{Entropy from Spacetime Correlation Functions}
\label{sec:SfromW}
Conventionally, entanglement entropy is defined to be 
\begin{equation}
    S = -\Tr(\rho_{red} \ln{\rho_{red}}),
\label{ee}
\end{equation}
where $\rho_{red}$ is the reduced density matrix after tracing out some degrees of freedom from an initially pure global state $\rho$. The degrees of freedom traced out lie within one of two causally complementary partitions of the spacetime (or in practice, Cauchy hypersurface).

The definition of entanglement entropy we work with is equivalent to \eref{ee}, but has been rewritten explicitly in terms of spacetime correlation functions. For a derivation and further details we refer the reader to \cite{Sorkin:2012sn}. It was shown in \cite{Sorkin:2012sn} that the entropy of a Gaussian scalar field in a spacetime region $\mathcal{R}$ is

\begin{equation}
    S = \sum_{\lambda} \lambda \log{|\lambda|},
    \label{S}
\end{equation}
where \(\lambda\) are the eigenvalues obtained by solving the generalised eigenvalue problem

\begin{equation}
    \mathbf{W}v = i \lambda \mathbf{\Delta} v,\indent \mathbf{\mathbf{\Delta}} v\neq 0.
    \label{geneig}
\end{equation}
\(i\mathbf{\Delta}\) is the Pauli-Jordan function, or spacetime commutator of the field operators, and \(\mathbf{W}\) is the Wightman, or two-point correlation function.
For a Gaussian scalar field theory,
\begin{equation}
    i\mathbf{\Delta}(x,y) := \langle0| [\Phi(x), \Phi(y)] |0\rangle=[\Phi(x), \Phi(y)],
\end{equation}
and
\begin{equation}
    \mathbf{W}(x,y) := \langle0| \Phi(x) \Phi(y) |0\rangle,
\end{equation}
where $x,y \in \mathcal{R}$. Note also the relations $ i\mathbf{\Delta}(x,y)=  \mathbf{W}(x,y)- \mathbf{W}(y,x)= \textrm{Im} (\mathbf{W}(x,y))/2$.
If we restrict \(\mathbf{\Delta}\) and \(\mathbf{W}\) to a subdomain of  $\mathcal{R}$ with nonempty causal complement, the entropy that is calculated via \eref{S} will correspond to the entanglement entropy between the subdomain and its complement. It is worth highlighting again that the formulas above are covariant and they solely involve spacetime correlation functions. 

The formulas \eref{S} and \eref{geneig} have been used in several settings both in the continuum theory and in causal set theory to compute entanglement entropies \cite{Saravani_2014, Sorkin_2018, Belenchia_2018, Surya_2021, Mathur_2021}. It has also been shown that these formulae capture the entanglement entropy even in non-Gaussian and interacting theories, up to first order in perturbation theory \cite{Chen_2020}. For theories where $ i\mathbf{\Delta}$ and $\mathbf{W}$ are known both in the continuum and in the causal set, the causal set calculations are easier to perform than the continuum ones. This is because $ i\mathbf{\Delta}$ and $\mathbf{W}$ are finite dimensional matrices and the generalised eigenvalue problem \eref{geneig} can be numerically solved relatively easily. In the continuum, $ i\mathbf{\Delta}$ and $\mathbf{W}$ are infinite dimensional, and must be made into finite dimensional matrices with respect to a choice of cutoff (see e.g. \cite{Saravani_2014}). This can be a challenging task in practice. 

In this work we focus on the causal set case, for practical ease, and also because it is more fundamental and equipped with a natural UV cutoff. In the causal set, the entanglement entropy is finite, and the degrees of freedom are finite and well-defined with respect to the discreteness scale, though as we will see, extra ``truncations'' are necessary to obtain meaningful results.

\section{Entanglement Entropy of a Single 1+1D Causal Interval}
\label{sec:oneint}

In order to use  \eref{S} and \eref{geneig} to study the entanglement entropy, we need a vacuum state, or Wightman function \(\mathbf{W}\). The only known means to define a Wightman function in causal set theory is via the Sorkin-Johnston (SJ) prescription \cite{Johnston_2008, johnston2010quantum, Afshordi2_2012, Sorkin:2017fcp}. This prescription is very much in the same spirit as the entropy formulation we reviewed in the previous section, as it  relies solely on spacetime quantities to define a vacuum state, and does not make reference to any spatial quantities on hypersurfaces. We review this prescription below.

\subsection{The SJ Prescription}
\begin{figure}[H]
    \centering
    \includegraphics[width=10cm]{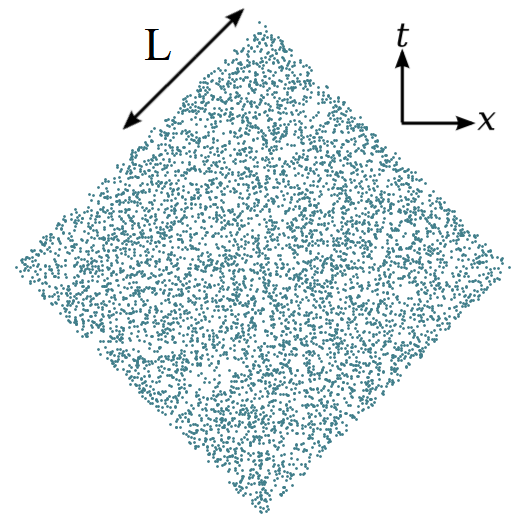}

    \caption{A causal set sprinkling of a causal diamond or interval in $1+1$D Minkowski spacetime.}
\label{diamond}
\end{figure}

The starting point of the SJ prescription is the retarded Green function. Since a unique retarded Green function exists only in globally hyperbolic spacetimes, the SJ vacuum is unique only in globally hyperbolic spacetimes. We will always work with causal set sprinklings\footnote{For a review of causal set theory we refer the reader to \cite{PhysRevLett.59.521, Surya_2019}.} approximated by $1+1$D causal intervals (diamonds) in Minkowski spacetime, as illustrated in Figure \ref{diamond}. These intervals are globally hyperbolic, and therefore have a unique retarded Green function and SJ state. We will work with a massless theory, for simplicity, and a Gaussian theory, since this is what the SJ prescription provides.

\FloatBarrier

The retarded Green function in $1+1$D Minkowski spacetime for a massless scalar field is 
\begin{equation}
    G_R(x, y) = \frac{1}{2}\Theta(\tau^2)\Theta(x^{0} - y^{0}),
\end{equation}
where \(\Theta\) is the Heaviside step function, and \(\tau\) is the proper time, defined via \(\tau^2=(x^0-y^0)^2-(x^i-y^i)^2\). This propagator is very simple, returning a value of one half if \(y\) is in \(x\)'s past lightcone. It is reminiscent of the causal matrix in the causal set, which is defined as

\begin{equation}
    \mathbf{C_{xy}} := \left\{ \begin{array}{ll}
      1 & x \prec y, \: \textrm{and} \:\, x \neq y\\
      0 & \textrm{otherwise}.
    \end{array} \right.
\end{equation}
We have used the notation that the indices $_{xy}$ designate the matrix element relating causal set elements $x$ and $y$, and $x \prec y$  denotes that $x$ causally precedes $y$. Indeed, the causal set retarded (and advanced) Green function, $\mathbf{K}_{R}$ (and $\mathbf{K}_{A}$), is defined precisely using the causal matrix:

\begin{equation}
    \mathbf{K}_{R} = \frac{1}{2}\mathbf{C}^T, \quad \textrm{and} \quad \mathbf{K}_{A} = \mathbf{K}_{R}^{T} = \frac{1}{2}\mathbf{C},
    \label{kra}
\end{equation}
where we have used the symbol $\mathbf{K}$ for the causal set Green function, to distinguish it from the continuum one $G$, and the superscript $^T$ refers to the transpose operation.

From $\mathbf{K}_{R,A}$ we can construct $\mathbf{\Delta}=\mathbf{K}_{R}-\mathbf{K}_{A}$. Because \(i\mathbf{\Delta}\) is Hermitian and antisymmetric, it can be expanded in its eigenbasis as

\begin{equation}
    i\mathbf{\Delta} = \sum_{i}^N\Lambda_{i}v_iv_i^{\dagger},
    \label{PJ}
\end{equation}
where $N$ is the number of elements in the causal set. \(\Lambda_{i}\) denotes an eigenvalue of \(i\mathbf{\Delta}\), with \(v_{i}\) being the corresponding eigenvector.

The SJ Wightman function, \(\mathbf{W}_{SJ}\) is defined to be the positive part of the expansion of \(i\mathbf{\Delta}\) in its eigenbasis in \eref{PJ}, i.e.

\begin{equation}
    \mathbf{W}_{SJ} := \sum_{i}\Lambda_{i}v_iv_i^{\dagger}, \quad \Lambda_i > 0.
    \label{CS_W}
\end{equation}
The SJ Wightman function defines a pure state for the entropy formulation we reviewed earlier, as it yields $\lambda=0 \textrm{ or } 1$ in \eref{geneig}, and hence a vanishing entropy in \eref{S}. Its restriction to a subregion with nonzero causal complement, however, will yield nontrivial $\lambda$'s, and a nonvanishing entanglement entropy.

For the spacetime interval in Figure \ref{diamond}, $W_{SJ}$ has been well studied both in the causal set and continuum \cite{Afshordi_2012}. Away from the boundaries of the diamond, which is where we will primarily place the subintervals we work with, the state resembles that of the Minkowski vacuum with an IR cutoff. Close to the left or right corner of the diamond, the state resembles that of the Minkowski vacuum in the presence of a static mirror or reflecting boundary at that corner.

\subsection{Truncations}
\label{sec:trunc}

For reasons which we will not delve into, it turns out that a stricter condition than $\mathbf{\Delta} v\neq 0$ in \eref{geneig} is necessary to obtain meaningful results in the causal set. Not only must we exclude the kernel of $i\mathbf{\Delta}$, we also need to exclude a large number of additional components corresponding to eigenvectors with nonzero but small magnitude eigenvalues \cite{Sorkin_2018}. To understand how we decide which components to keep and which to exclude, it is instructive to first review some aspects of $i\mathbf{\Delta}$ and its eigendecomposition in Minkowski spacetime.

In lightcone coordinates \(u=\frac{x+t}{\sqrt{2}}\) and \(v=\frac{t-x}{\sqrt{2}}\), the Pauli-Jordan function in the continuum theory is \cite{Afshordi_2012} 
\begin{equation}
    i\mathbf{\Delta}(u,v;u',v') = \frac{-i}{2}(\Theta(u-u') + \Theta(v-v') - 1).
\end{equation}

Its eigenfunctions $f$ satisfy 
\begin{equation}
     \int_{\mathcal{R}} i\mathbf{\Delta}(u,v;u',v') f(u,v) \: du\, dv = \Lambda f(u',v'),
     \label{ceig}
\end{equation}
where the domain of integration, $\mathcal{R}$, is the causal diamond with origin at the center and $u\in[-L,L]$ and $v\in[-L,L]$. This causal interval is the continuum approximation to the causal set in Figure \ref{diamond}.

The eigenfunctions with non-zero eigenvalues\footnote{It is worth noting that this eigenspace spans the solution space of the Klein-Gordon equation.} satisfying \eref{ceig} are \cite{johnston2010quantum, Afshordi_2012}
\begin{equation}
    f_k(u, v) = e^{-iku} - e^{-ikv}, \qquad k = \frac{n\pi}{L}, \: n \in \mathbb{Z}\backslash 0
    \label{single_diamond_soln1}
\end{equation}
and
\begin{equation}
    \begin{array}{ll}
    & g_k(u, v) = e^{-iku} + e^{-ikv} - 2\cos{(kL)}, \\
    & k \in \mathcal{K}= \{k \in \mathbb{R} | \tan{(kL)}=2kL \: \textrm{and} \: k \neq 0 \}.
    \end{array}
    \label{single_diamond_soln2}
\end{equation}
Each \(f_k(u, v)\) and \(g_k(u, v)\) has corresponding eigenvalue
\begin{equation}
    \Lambda_k =\frac L k,    
\end{equation}
and for large \(k\), the \(g_k\) eigenvalues tend to those of \(f_k\), i.e. $\Lambda_k = \frac{L^2}{n \pi}$. When calculating entanglement entropies in the continuum via the definition in Section \ref{sec:SfromW}, a minimum eigenvalue serving as the UV cutoff is arbitrarily set \cite{Saravani_2014}, given by \(\Lambda_{min} = \frac{L^2}{n_{max}\pi}\). Via dimensional considerations, and a direct comparison of values, it is seen that the eigenvalues in the causal set are related to those in the continuum by rescaling with a factor of the density \(\rho = \frac{N}{4L^{2}}\), such that

\begin{equation}
    \Lambda^{cs} = \rho \Lambda^{cont}.
    \label{lamscale}
\end{equation}

In the causal set, the value of the cutoff and minimum eigenvalue are not arbitrary. The cutoff is the discreteness scale, and the minimum eigenvalue is related to it. Given the nature of the eigenfunctions as approximate linear combinations of plane waves and the eigenvalues as inversely proportional to wavenumbers, we expect a maximum wavenumber, \(k_{max}\), near the discreteness scale. Larger wavenumbers or shorter wavelengths cannot be meaningfully supported on the causal set. This upper bound on the wavenumber leads via \eref{lamscale} to a minimum (in magnitude) expected eigenvalue given by

\begin{equation}
    \Lambda^{cs}_{min} =\frac {\rho L} {k_{max}}=\frac{\sqrt{N}}{4\pi},
\end{equation}
where $N$ is the number of elements in the diamond in which the eigenvalue problem is solved. As first noted in \cite{Sorkin_2018}, there are residual contributions below this $\Lambda^{cs}_{min}$ in the spectrum of $i\mathbf{\Delta}$ in the causal set. These need to be removed through \emph{truncations}.  The truncations are a procedure that take out or set to zero any components that correspond to eigenfunctions with eigenvalues smaller (in magnitude) than $\Lambda^{cs}_{min}$. The truncations need to be done at two stages of the calculation: during the SJ prescription, and after the restriction to the entangling subregion but before solving \eref{geneig}. For further details on the truncation process, we refer the reader to \cite{Sorkin_2018, YKY}. 

\subsection{Subinterval in a Corner}
\label{sec:corner}
\begin{figure}[h!] 
    \centering
    \includegraphics[width=10cm]{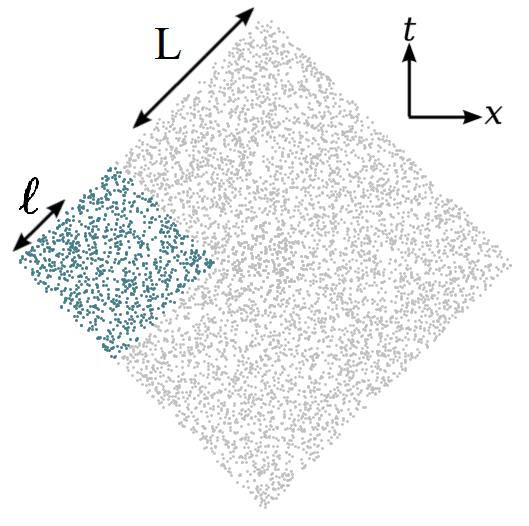}

    \caption{A causal set subinterval in the left corner of a larger causal set interval.}
\label{corner}
\end{figure}

We now use everything we have reviewed above to calculate the entanglement entropy of the restriction of a massless scalar field to a causal set subinterval in the left corner of a larger one (such that they share a boundary). The causal intervals are shown in Figure \ref{corner}. Our calculation is similar to that in \cite{Sorkin_2018}, the only difference being that the subinterval considered there was concentric to the larger one and had two boundaries across which there was entanglement. In our setup there is only one boundary contributing to the entanglement entropy.
\FloatBarrier
For the one boundary setup, in the limit $\ell\ll L$ and the UV cutoff $a\rightarrow 0$, the expected entanglement entropy scaling with the UV cutoff is \cite{Calabrese_2004}
\begin{equation}
    S_{corner} = \frac{1}{6} \ln\left(\frac{\ell}{a}\right) + b,
    \label{1bndryS}
\end{equation}
where $b$ is a non-universal constant. Note that a logarithmic scaling with the UV cutoff in $1+1$D is consistent with the spatial area law of entanglement entropy \cite{Chandran_2016}.

Our choice of UV cutoff is the causal set discreteness scale $a\equiv1/\sqrt{\rho}=2L/\sqrt{N_L}=2\ell/\sqrt{N_\ell}$. In our calculations we hold fixed the volumes of the spacetime intervals $L=50$ and $\ell/L=2/5$ and vary $a$ by changing the number of sprinkled elements $N_L$ in the larger diamond.

\begin{figure}[h!]
    \centering
    \includegraphics[width=16cm]{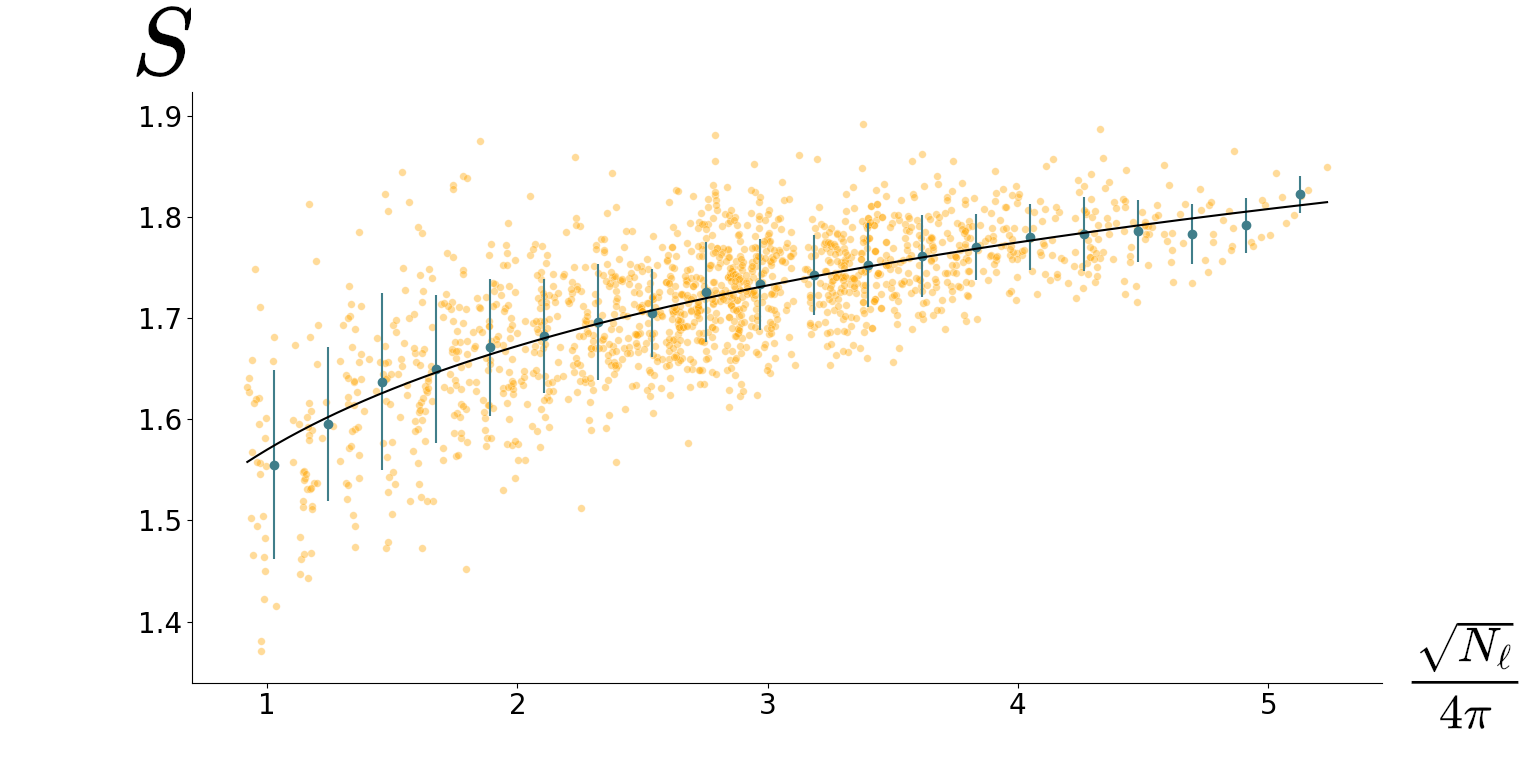}

    \caption{The scaling of entanglement entropy with respect to the UV cutoff, for the corner diamond configuration. The raw data has been plotted (orange), as well as averaged data points where binning has been performed (blue). The error bars give the standard deviation of the bins (blue). A function of the form \(\alpha\ln(x)+\beta\), shown in black, was fit to the binned points with coefficient \(\alpha = 0.165 \pm 0.0195\), and \(\beta = 1.553 \pm 0.0278\).}
\label{side_diamond_S}
\end{figure}

Our results for the entanglement entropy, computed through \eref{geneig} and \eref{S}, versus $\frac{\sqrt{N_\ell}}{4\pi}$ are shown in Figure \ref{side_diamond_S}. $\frac{\sqrt{N_\ell}}{4\pi}$ is the minimum eigenvalue of $i\mathbf{\Delta}$ in the subinterval and is proportional to $1/a$. Included in the figure is a best fit logarithmic scaling which is

\begin{equation} 
    S = (0.165 \pm 0.0195)\ln{\left(\frac{\sqrt{N_\ell}}{4\pi}\right)} + (1.553 \pm 0.0278),
\end{equation}
in good agreement with the expected scaling and coefficient in \eref{1bndryS}.

\FloatBarrier

\section{Entanglement Entropy of Two 1+1D Disjoint Intervals}
\label{sec:2int}
The entanglement entropy associated with multiple disjoint intervals has been a topic of great interest \cite{Arias_2018, Alba_2010, Casini_2004, Calabrese:2010he, Ruggiero_2018, PhysRevD.100.106015, hartman2013entanglement}. A feature which makes these intervals interesting to study in conformal field theory, but also very difficult to perform calculations with, is that the entanglement entropy depends on the full operator content of the theory \cite{Alba_2010}. This is in contrast to the entanglement entropy of a single interval which only depends on the central charge. Disjoint intervals also allow for the study of more complex entanglement properties, as well as the study of mutual information and relative entropy. Relative entropy is argued by some to be a better measure of microscopic degrees of freedom than entanglement entropy, for reasons including that it is finite in the UV or continuum limit \cite{hollands2018entanglement, Witten_2018, swingle2010mutual}. We will study the mutual information in Section \ref{sec:mutualinfo}. 

\subsection{Entanglement Entropy}
\begin{figure}[h!]
    \centering
    \includegraphics[width=10cm]{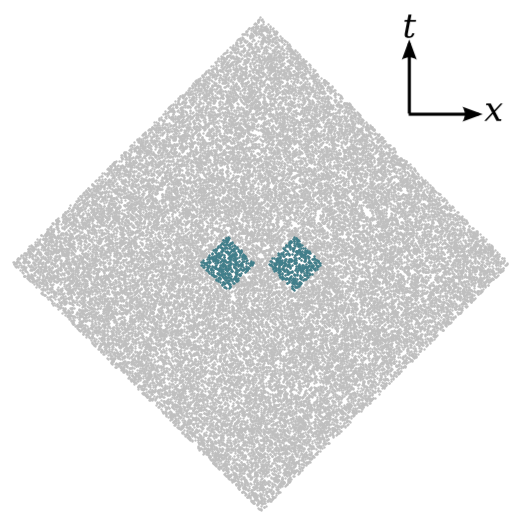}

    \caption{Two disjoint subdiamonds within a larger causal set interval. }
\label{two_intervals}
\end{figure}

In this subsection, we carry out a similar calculation to the one in Section \ref{sec:corner}, but considering a subregion comprised of two disjoint causal diamonds (with the same volume) within a larger single diamond that does not share any boundaries with the subintervals. The disjoint causal intervals are shown in Figure \ref{two_intervals}. We place the subdiamonds away from the boundaries of the larger diamond, such that they are in the regime where the SJ state resembles that of the Minkowski vacuum with an IR cutoff. For two intervals, in the limit $\ell\ll L$ and  $a\rightarrow 0$, the  entanglement entropy scaling with $a$ is expected to be \cite{Casini_2004}
\begin{equation}
    S_{2int} = \frac{2}{3} \ln\left(\frac{\ell}{a}\right) + b',
    \label{2intS}
\end{equation}
where $b'$ is once again a non-universal constant.
\FloatBarrier
Since the global region is a single causal interval, as before, the SJ prescription and the first truncation go through identically to the previous section. Our second truncation, however, is different. This is because we are restricting to a different domain, namely that of two disjoint diamonds. Though the domain and eigenspace are different in this case, knowledge of the single diamond case is all that is required to generalise everything to two (and more) intervals. 

Let us once again return to the continuum theory, to motivate the second truncation we implement. Since the eigenspace of $i\mathbf{\Delta}$ with non-zero eigenvalues in each domain dictates which contributions we should keep, let us consider the eigenvalue problem \eref{ceig} in the union of the disjoint diamonds

\begin{equation}
    -\frac{i}{2} \int_{\mathcal{R}_2} (\Theta(u-u') + \Theta(v-v') - 1) f(u,v) du\,dv = \Lambda f(u',v'),
     \label{int2int}
\end{equation}
where the domain of integration, $\mathcal{R}_2\equiv \Diamond_1\cup \Diamond_2$, is now a union of two causally disjoint diamonds. We can also define this region in lightcone coordinates as

\begin{equation}
    \mathcal{R}_2 \equiv
    \begin{array}{ccc}
    -(2\ell+d) < u < -d & \textrm{ and } & d < v < 2 \ell +d \\ & \cup & \\ d < u < 2\ell + d & \textrm{ and } & -(2\ell+d) < v < -d,
    \end{array}
\end{equation}
where $d$ is the half diagonal separation. It is equal to half the shortest vertex to vertex separation between the subdiamonds, divided by $\sqrt{2}$. Since $i\mathbf{\Delta}$ is the spacetime commutator, it vanishes at pairs of points or elements that are spacelike to one another. The two causal intervals are spacelike to one another, and thus the field at every point in one diamond commutes with the field at every point in the other diamond. We can then see that $i\mathbf{\Delta}$ is block diagonal in the position basis, with a block for each disjoint interval. 

With this in mind, we can break up the integral \eref{int2int} into a sum of two integrals, each over one of the subdomains $\Diamond_1$ and $\Diamond_2$. The eigenfunctions with non-zero eigenvalues can be constructed as piecewise functions, analogous to the single diamond eigenfunctions \eref{single_diamond_soln1} and \eref{single_diamond_soln2}, with support on one causal interval at a time\footnote{If there are degenerate eigenvalues, any linear combination of their respective eigenfunctions is also a valid eigenfunction. Therefore, if there was an eigenfunction in $\Diamond_1$ and another one in $\Diamond_2$ with the same eigenvalue, one could take a linear combination of them which would be non-zero in both diamonds. In the causal set, because the two diamonds are not identical due to Poisson fluctuations, an exact degeneracy is much less likely, so this makes the eigenfunctions have support on only one diamond at a time.}:  

\begin{equation}
    f_k(u, v) = \left\{ \begin{array}{ll}
    e^{-iku} - e^{-ikv} & \{u,v\} \in \Diamond_1 \: \textrm{or} \: \Diamond_2 \\
    0 & \textrm{otherwise,}
    \end{array} \right.
    \label{disjoint_diamond_soln1}
\end{equation}
\begin{equation}
    k = \frac{n\pi}{\ell}, \quad n \in \mathbb{Z}\backslash 0, \nonumber
\end{equation}
and
\begin{equation}
    g_k(u, v) = \left\{ \begin{array}{ll}
    e^{-iku} + e^{-ikv} - 2\cos{(k\ell)} & \{u,v\} \in \Diamond_1 \: \textrm{or} \: \Diamond_2 \\
    0 & \textrm{otherwise,}
    \end{array} \right.
    \label{disjoint_diamond_soln2}
\end{equation}
\begin{equation}
        k \in \mathcal{K}= \{k \in \mathbb{R} | \tan{(k\ell)}=2k\ell \: \textrm{and} \: k \neq 0 \},
        \nonumber
\end{equation}
where for simplicity we have expressed each function in the coordinate system with its origin at the center of the diamond in which it has support. The eigenvalues are $\Lambda=\ell/k$, as before. We also empirically verified in the causal set that the eigenfunctions have support on one interval at a time and are approximately plane wave-like. That the eigenfunctions \eref{disjoint_diamond_soln1} and \eref{disjoint_diamond_soln2} are the full set of eigenfunctions follows directly from the single diamond assertion of this. 

Knowing the form of the eigenfunctions now enables us to define a truncation scheme for the disjoint diamonds. We can follow the same reasoning as in Section \ref{sec:trunc} to conclude that the minimum eigenvalue magnitude in each interval $i$ is
\begin{equation}
    \Lambda_{min,i}^{cs} = \frac{\sqrt{N_{i}}}{4\pi},
\end{equation}
where $N_i$ is the number of elements in the region $i$. For simplicity, we will consider the case where the disjoint diamonds have the same volume, such that $N_1\simeq N_2$\footnote{For the case of unequal volumes, the truncations must be performed separately in each disjoint region. Having equal volumes facilitates implementing the truncation, as the same minimum eigenvalue magnitude is valid over the entire union region.}, and for ease of calculation and presentation, we define $N_\ell := \langle N_i \rangle \simeq N_i$.

Once again our UV cutoff is the causal set discreteness scale $a\equiv1/\sqrt{\rho}$. In our calculations we hold fixed the volumes of the spacetime intervals $L=90$ and $\ell_1=\ell_2=10$, and the diamond half diagonal separation $d=2.2$. We vary $a$ by varying the number of sprinkled elements $N_L$ in the larger diamond.

\begin{figure}[h!]
    \centering
    \includegraphics[width=16cm]{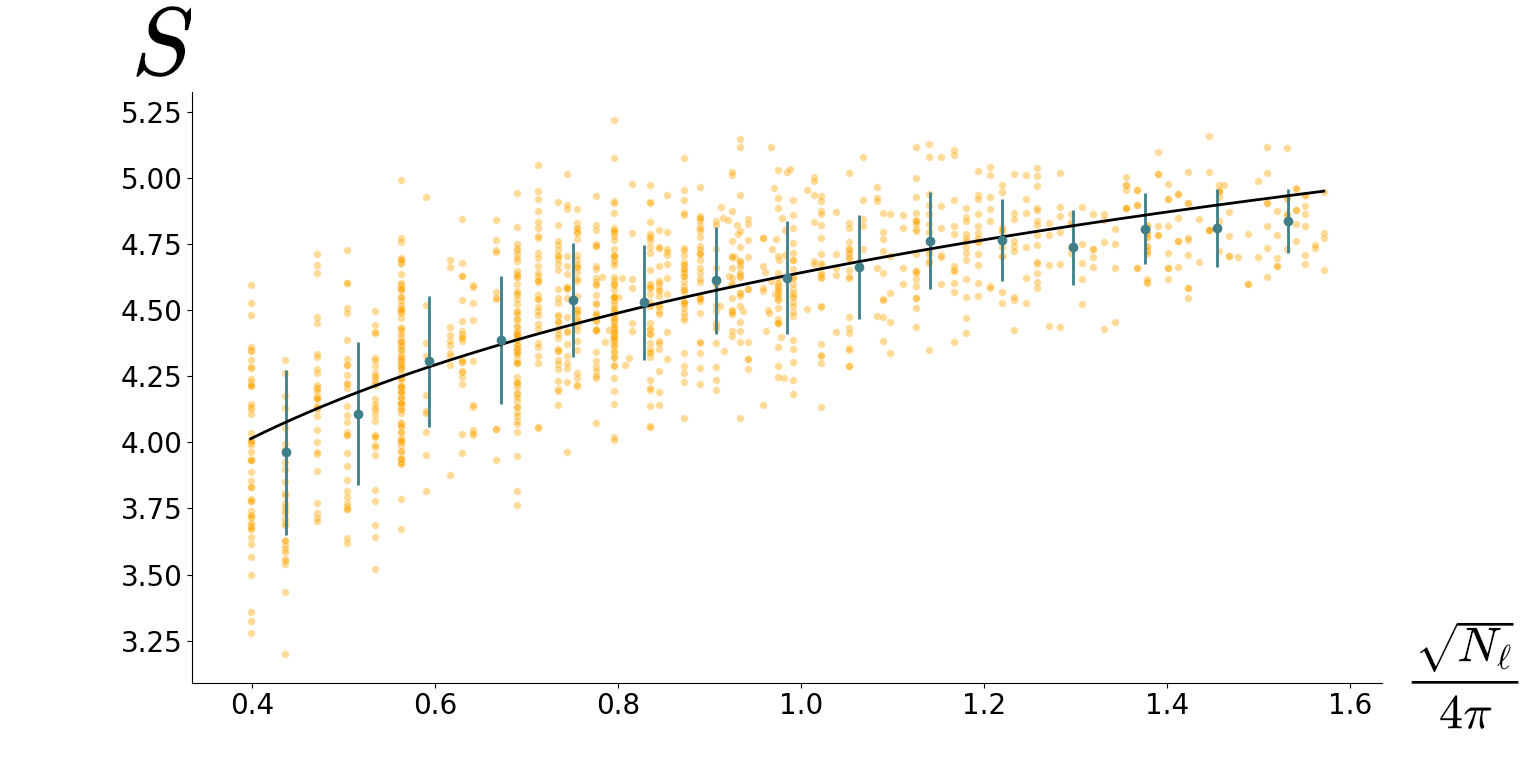}

    \caption{The scaling of entanglement entropy with respect to the UV cutoff, for the two disjoint subdiamond configuration. The raw data has been plotted (orange), as well as averaged data points where binning has been performed (blue). The error bars give the standard deviation of the bins (blue). A function of the form \(\alpha\ln(x)+\beta\), shown in black, was fit to the binned points with coefficient \(\alpha = 0.669 \pm 0.0207\), and \(\beta = 4.642 \pm 0.0077\).}
\label{two_diamond_S}
\end{figure}

Our results for the entanglement entropy, computed through \eref{geneig} and \eref{S}, versus $\frac{\sqrt{N_\ell}}{4\pi}$ are shown in Figure \ref{two_diamond_S}. Included in the figure is a best fit logarithmic scaling which is

\begin{equation}
    S = (0.669 \pm 0.0207)\ln{\frac{\sqrt{N_\ell}}{4\pi}} + (4.642 \pm 0.0077),
\end{equation}
in good agreement with the expected scaling and coefficient in \eref{2intS}.

\FloatBarrier

\subsection{Mutual Information}
\label{sec:mutualinfo}
Now that we can compute the entanglement entropy for single intervals as well as unions of two intervals, we can study the mutual information. The mutual information for disjoint  regions $A$ and $B$ is defined as \cite{Witten_2020}

\begin{equation}
    I_{A:B} := S_{A \cup B} - S_{A} - S_{B}.
\end{equation}
It is a non-negative quantity and for disjoint intervals such as the diamonds we considered above, the mutual information is expected to decay as the separation between the intervals increases \cite{Furukawa_2009}. This is intuitive as the field degrees of freedom become more and more uncorrelated as the spacelike separation between them increases. We will confirm this decay below.

\begin{figure}[h!]
    \centering
    \includegraphics[width=10cm]{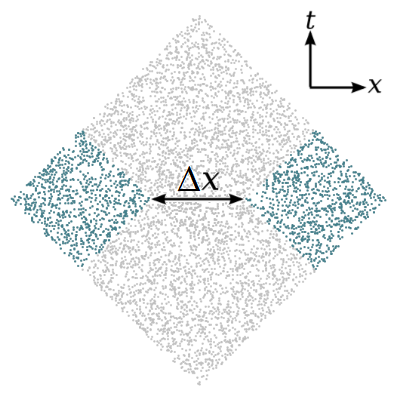}

    \caption{A figure showing the two disjoint diamonds configuration where we investigate the  mutual information.}
\label{two_interval_side}
\end{figure}

The causal set setup we consider is shown in Figure \ref{two_interval_side}. We place the subdiamonds at the extreme left and right corners of the larger diamond, to maximize the separation between them. We hold fixed the sizes of the subintervals, $\ell_1=\ell_2=5$, as well as the sprinkling density $\rho=10$. We then increase the separation between the two intervals by keeping the subintervals in the left and right corners of the larger diamond while increasing the size of the larger global diamond.

\FloatBarrier

\begin{figure}[h!]
    \centering
    \includegraphics[width=15cm]{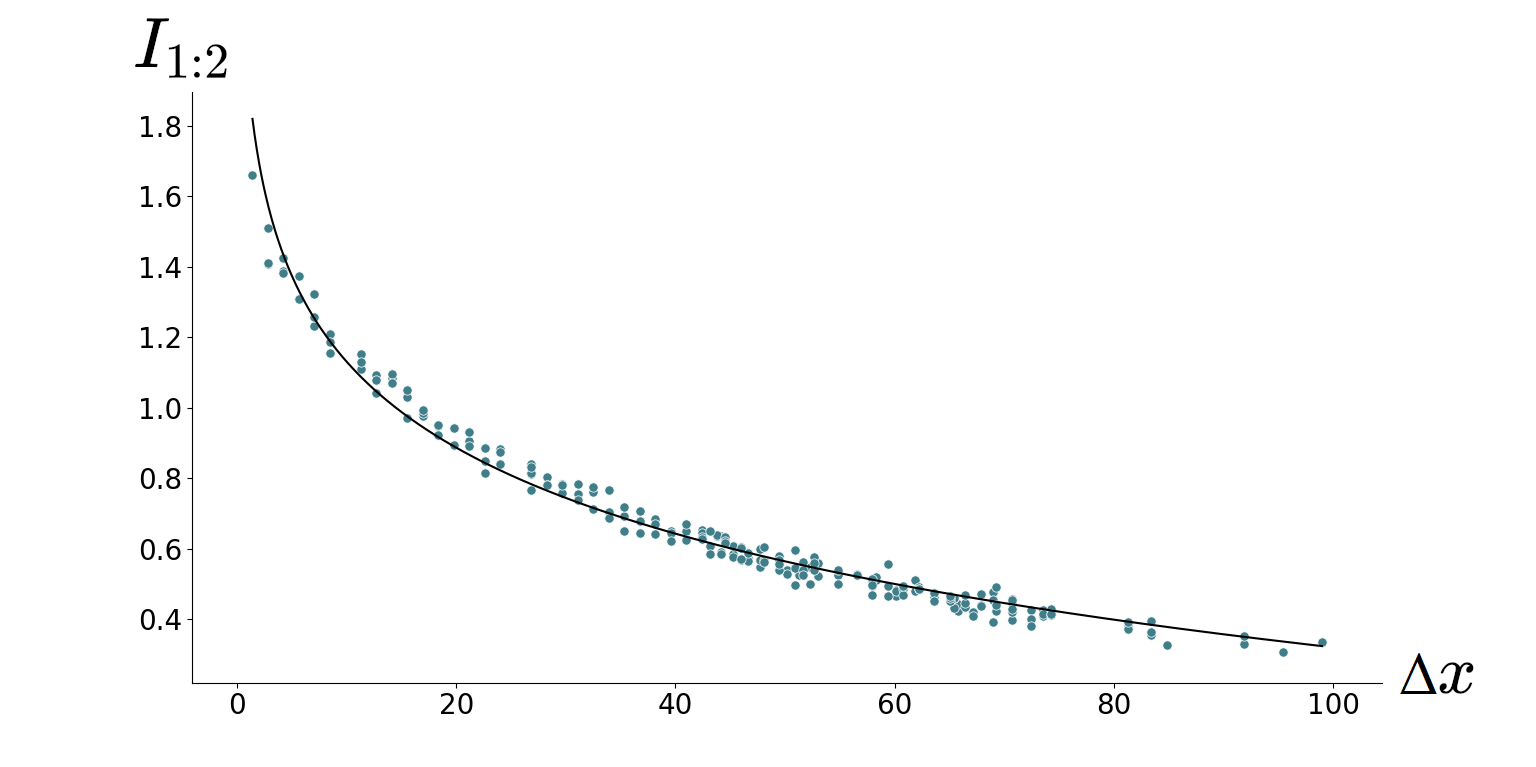}

    \caption{A figure showing the decay of mutual information as two disjoint diamonds become far separated from one another relative to their own sizes. The vertical axis gives the value for mutual information and the horizontal axis \(\Delta x\) is the separation between the two regions.}
\label{mutual_info}
\end{figure}

The result for the scaling of the mutual information $I_{1:2}$ versus the separation between the two intervals, $\Delta x$, measured using the horizontal separation between the innermost corners, is shown in Figure \ref{mutual_info}. As expected, the mutual information is non-negative and decreases as the separation $\Delta x$ increases.   
The decay trend is approximated by 

\begin{equation}
 I_{1:2} = \alpha \log \left(\frac{\Delta x}{\Delta x+30}\right) + \beta\left(\frac{\Delta x}{\Delta x+30} -1\right),
 \label{mi_fit}
\end{equation}
with best fit parameters \(\alpha = -0.0926 \pm  0.00797\) and \(\beta = -1.3783 \pm  0.01342\). The reason $30$ appears in \eref{mi_fit} is that this is approximately the sum of the diameters of the two subdiamonds, and how far they are from one another should be considered with respect to this scale. The curve \eref{mi_fit} is shown together with the data in Figure \ref{mutual_info}. Once again, as anticipated,  $I_{1:2}\rightarrow0$ as $\Delta x\rightarrow\infty$ in \eref{mi_fit}. Also, as expected, there is a divergence as the two disjoint regions approach one another \cite{swingle2010mutual}. %This plot verifies the decay of mutual information with respect to the diamond separation tends to zero for large separations and diverges as the separation decreases, but more investigation should be done into this functional form. There has not been an investigation into an expression for mutual information in the presence of boundaries in the conformal field theory case, and such a scaling, if found, would be interesting to verify in the causal set, as more concrete evidence of the veracity of our truncation scheme.

\FloatBarrier

\section{Entanglement Entropy of Three 1+1D Disjoint Intervals}
\label{sec:3int}
\begin{figure}[h!]
    \centering
    \includegraphics[width=10cm]{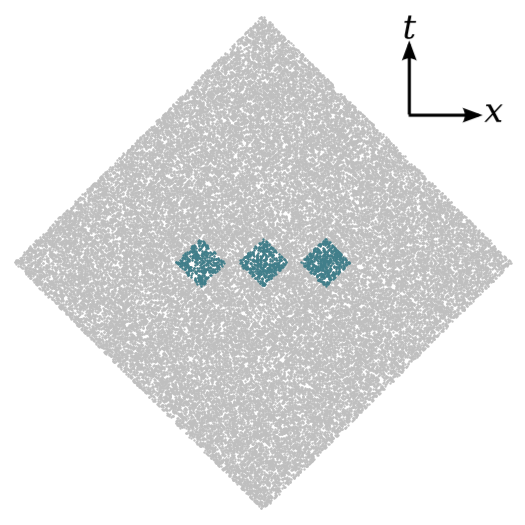}

    \caption{Three disjoint subdiamonds within a larger causal set interval.}
\label{three_intervals}
\end{figure}

As a final example, we consider the entanglement entropy of three disjoint causal intervals. The causal set setup is illustrated in Figure \ref{three_intervals}. Once again for ease of applying the truncations, we set the volumes of the three diamonds to be equal, though more general configurations can be considered as well.
Similar to the case of two diamonds in Section \ref{sec:2int}, none of the subdiamonds share any boundaries with the larger diamond they lie within, and they reside in the region where the SJ state resembles the Minkowski vacuum with an IR cutoff. 

\FloatBarrier

For three intervals, in the limit $\ell\ll L$ and  $a\rightarrow 0$, the  entanglement entropy scaling with $a$ is expected to be 
\begin{equation}
    S_{3int} =  \ln\left(\frac{\ell}{a}\right) + b'',
    \label{3intS}
\end{equation}
where $b''$ is again a non-universal constant.

As above, we hold fixed the sizes of the three intervals, $\ell_1=\ell_2=\ell_3=5$, as well as the size of the larger diamond, $L=50$. We vary the sprinkling density (by changing $N$) to vary $a$. The first truncation is the same as before, and in the second truncation, the minimum magnitude eigenvalue of $i\mathbf{\Delta}$ in each interval $i$ is
\begin{equation}
    \Lambda_{min,i}^{cs} = \frac{\sqrt{N_{i}}}{4\pi},
\end{equation}
where $N_i$ is the number of elements in the region $i$. 

\begin{figure}[h!]
    \centering
    \includegraphics[width=16cm]{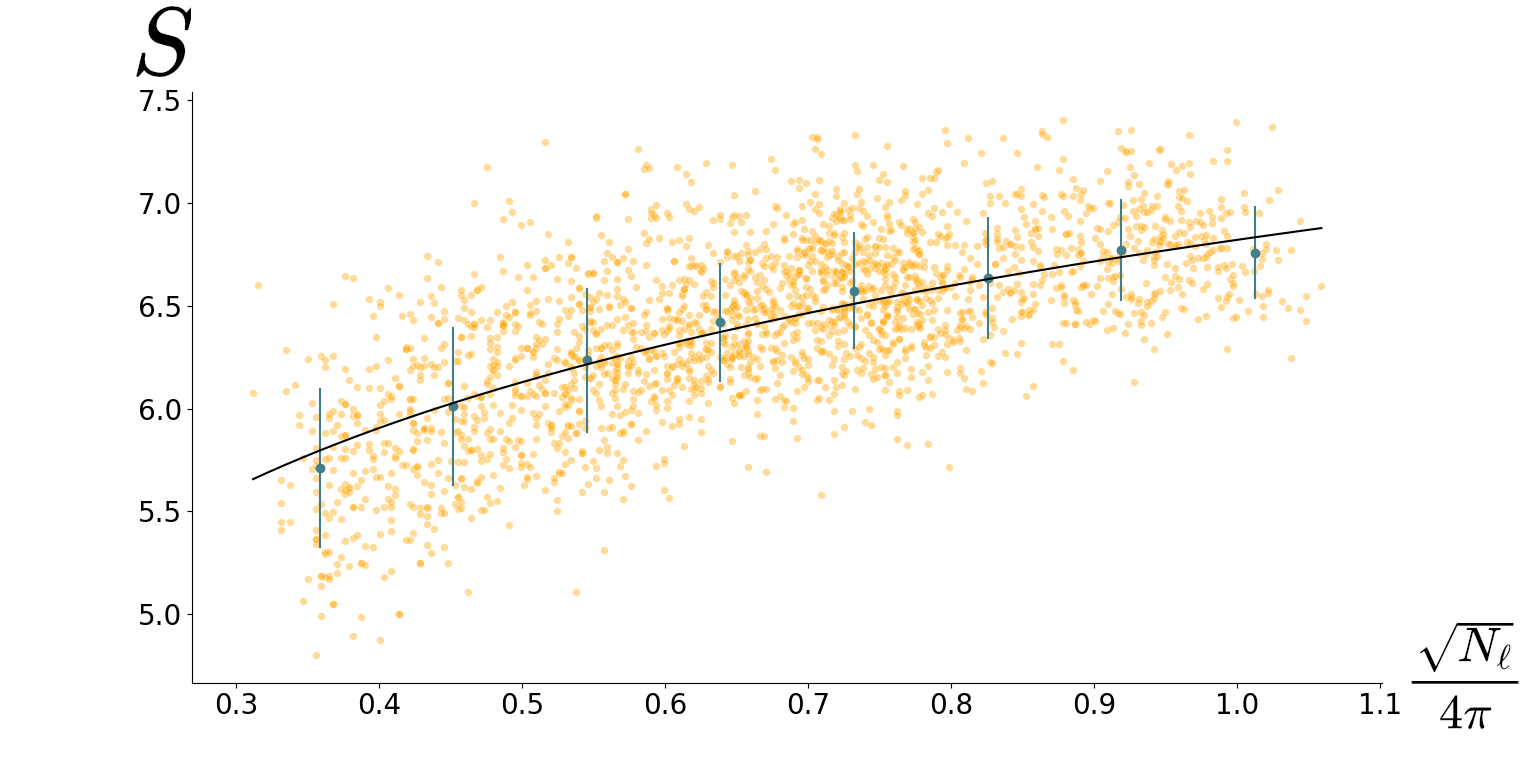}

    \caption{The scaling of entanglement entropy with respect to the UV cutoff, for the three disjoint subdiamond configuration. The raw data has been plotted (orange), as well as averaged data points where binning has been performed (blue). The error bars give the standard deviation of the bins (blue). A function of the form \(\alpha\ln(x)+\beta\), shown in black, was fit to the binned points with coefficient \(\alpha = 0.998 \pm 0.0691\), and \(\beta = 6.821 \pm 0.0307\).}
\label{three_diamond_S}
\end{figure}

Our results for the entanglement entropy, computed through \eref{geneig} and \eref{S}, versus $\frac{\sqrt{N_\ell}}{4\pi}$ are shown in Figure \ref{three_diamond_S}. Included in the figure is a best fit logarithmic scaling which is
\begin{equation}
    S = (0.998 \pm 0.0691)\ln{\frac{\sqrt{N_\ell}}{4\pi}} + (6.821 \pm 0.0307),
\end{equation}
yet again in good agreement with the expected coefficient of \(1\) on the logarithm as in \eref{3intS}.

\FloatBarrier

\section{Conclusions and Future Directions}
In this paper we have studied the entanglement entropy of a scalar field in disjoint intervals or causal diamonds in causal set theory. We used a spacetime definition of entanglement entropy in terms of correlation functions and found that it is especially suited to calculations involving disjoint causal regions. Not only does the definition admit a covariant UV cutoff, it is also applicable in causal set theory, where conventional hypersurface-based calculations cannot be used. One reason that the spacetime formalism we use is natural for disjoint regions is that the spacetime commutator plays a central role in it. The commutator is only nonzero at pairs of points that are causally related, and this simplifies things considerably. The subtleties that need to be taken care of (e.g. truncations in the causal set case) in the multiple interval cases boil down to the same ones that are present in the single interval case. In two spacetime dimensions, the single interval case is well studied, facilitating the multiple interval generalisations.

We specifically considered the cases of two and three disjoint causal intervals. We studied the entanglement entropy scaling with respect to the UV cutoff, which is the discreteness scale in the causal set, and found agreement with the expected area law scalings from other similar studies. We also considered the mutual information of two disjoint intervals and verified that it decays to zero as we increase the separation between the two intervals. 

There are many extensions of our work that would be interesting to explore in future research. As mentioned, we primarily focused on the entanglement entropy scalings with the UV cutoff. One can also study the scalings with respect to the other length scales in the problem, namely the sizes of each of the diamonds (including both the global one and the subdiamonds). These scalings in the scalar theory have in prior work  been analytically challenging to derive. Our work presents a framework where the numerical study of this is possible. As we have mentioned, the formalism we use is suited for the study of multiple intervals, as much of what we know from the single interval case carries over. This is true for higher dimensions as well, and it would be interesting to extend these results to higher dimensions. In order to do this, first the single interval case in higher dimensions needs to be better studied. The causal set entanglement entropy calculations could be a useful tool in other fields as well, as they make otherwise challenging calculations comparably easier to perform, as demonstrated by our examples. 

Our techniques could also be used to study the nature of the non-universal constant (the $b$’s or $\beta$’s above) in the entanglement entropy. There is a large literature on entanglement entropy in topological quantum field theories (e.g. \cite{Kitaev_2006, PhysRevLett.96.110405, tqft1}). In these works, the entanglement entropy separates into a contribution that scales with the boundary of the subsystem and a remaining contribution that does not. The latter is called the topological entanglement entropy, and it can be isolated via the addition and subtraction of entropies of subdivisions of the region of interest. In the same manner, perhaps similar manipulations can be implemented in the context of our work, to isolate some (topological or otherwise) properties of the non-universal constant.

Finally, we have focused on entanglement entropies in the context of causal set theory. The causal set calculations are more straightforward than their continuum counterparts, and they are also more fundamental. However, the formalism we use is equally applicable in the continuum theory. It would be interesting to investigate entanglement entropies of multiple intervals in the continuum, in the same manner as we have done above in the causal set. This would offer the possibility of deriving analytic results in this context.
\newline

\bf Acknowledgements: \rm
We thank Horacio Casini and Fay Dowker for helpful discussions. YY acknowledges financial support from Imperial College London through an Imperial College Research Fellowship grant.

\section*{References}
\bibliographystyle{iopart-num-long}
\bibliography{references.bib}

\end{document}